\documentclass{ctr_summer}

\usepackage{ctrfont}
\usepackage{natbib}
\usepackage{undertilde}
\usepackage{color}
\usepackage{bm}
\usepackage{subcaption}
\renewcommand{\figurename}{\textsc{Figure}}
\renewcommand{\tablename}{\textsc{Table}}
\usepackage{mathtools}
\usepackage{mathrsfs}
\usepackage{amsfonts}

\usepackage{graphicx}

\usepackage{url}

\usepackage{color}

\definecolor{deepred}{rgb}{0.,0,0.6}

\newcommand{\eps}{\varepsilon}

\newcommand\mapsfrom{\mathrel{\reflectbox{\ensuremath{\mapsto}}}}
\newcommand*\mean[1]{\bar{#1}}
\newcommand{\datacov}{\mathcal{C}_{\bm{u}^\star}}
\newcommand{\priorcov}{\mathcal{C}_{\mean{\bm{x}}}}
\newcommand{\postcovmap}{\mathcal{C}_{\bm{x}^\circ}}
\newcommand{\postcovxk}{\mathcal{C}_{\bm{x}_k}}

\DeclarePairedDelimiter{\abs}{\lvert}{\rvert}
\DeclarePairedDelimiter{\norm}{\lVert}{\rVert}

\DeclarePairedDelimiter{\binner}{\big\langle}{\big\rangle}

\DeclarePairedDelimiterX{\dotp}[2]{\big\langle}{\big\rangle}{#1, #2}
\DeclarePairedDelimiterX{\Dotp}[2]{\Big\langle}{\Big\rangle}{#1, #2}

\title{Bayesian inference of mean velocity fields and
turbulence models from flow MRI}

\shorttitle{Bayesian inference of mean velocity fields from flow MRI}

\author{A. Kontogiannis\footnote[1]{Department of Engineering, University of Cambridge, United Kingdom}, P. Nair, M. Loecher, D. B. Ennis, A. Marsden \and \mbox{M. P. Juniper}$\dagger$}

\shortauthor{Kontogiannis et~al.}

\begin{document}

\setcounter{page}{1}
\maketitle

We solve a Bayesian inverse Reynolds-averaged Navier--Stokes (RANS) problem that assimilates mean flow data by jointly reconstructing the mean flow field and learning its unknown RANS parameters. We devise an algorithm that learns the most likely parameters of an algebraic effective viscosity model, and estimates their uncertainties, from mean flow data of a turbulent flow. We conduct a flow MRI experiment to obtain mean flow data of a confined turbulent jet in an idealized medical device known as the FDA (Food and Drug Administration) nozzle. The algorithm successfully reconstructs the mean flow field and learns the most likely turbulence model parameters without overfitting. The methodology accepts any turbulence model, be it algebraic (explicit) or multi-equation (implicit), as long as the model is differentiable, and naturally extends to unsteady turbulent flows.\\

\hrule

\section{Introduction}
\par
Probability theory \citep{Cox1946,Jaynes2003} provides a fundamental and rigorous framework for extracting information from data when we know, to some degree of certainty, what the data contain. In this approach, some parameters are fixed while others are expressed as probability distributions \citep{MacKay2003,Stuart2010,Ghattas2021}. These distributions are initially asserted by the user and are then updated deterministically with Bayesian inference as data arrive. If more than one model could explain the data, the likelihood of each model is calculated and the most likely model is selected \citep{MacKay2003}. This framework provides a principled way to tune and choose models, given the available data. This is also known as inverse problem theory, described in seminal texts such as \citet{Kaipio2005,Idier2002,Sullivan2015}.

The challenge with performing Bayesian inference on partial differential equations (PDEs) in general and fluid dynamics in particular is that function evaluations are usually expensive. This means that sampling methods such as Markov chain Monte Carlo (MCMC) or Hamiltonian Monte Carlo (HMC) quickly become intractable \citep{Kaipio2005,Sullivan2015}. For problems in which the outputs vary smoothly with their parameters, however, we can exploit the fact that Gaussian prior distributions lead to nearly Gaussian posterior distributions, meaning that Laplace’s approximation (a second-order Taylor expansion of the model output with respect to its parameters) becomes sufficiently accurate to be useful \citep{MacKay2003,Sullivan2015}. Laplace’s approximation can then be combined with adjoint methods such that many parameters of large PDE problems can be inferred simultaneously and cheaply. 

\begin{figure}
    \centering
    \includegraphics[trim = 1cm 8cm 1.7cm 4cm, clip=true, width=\linewidth]{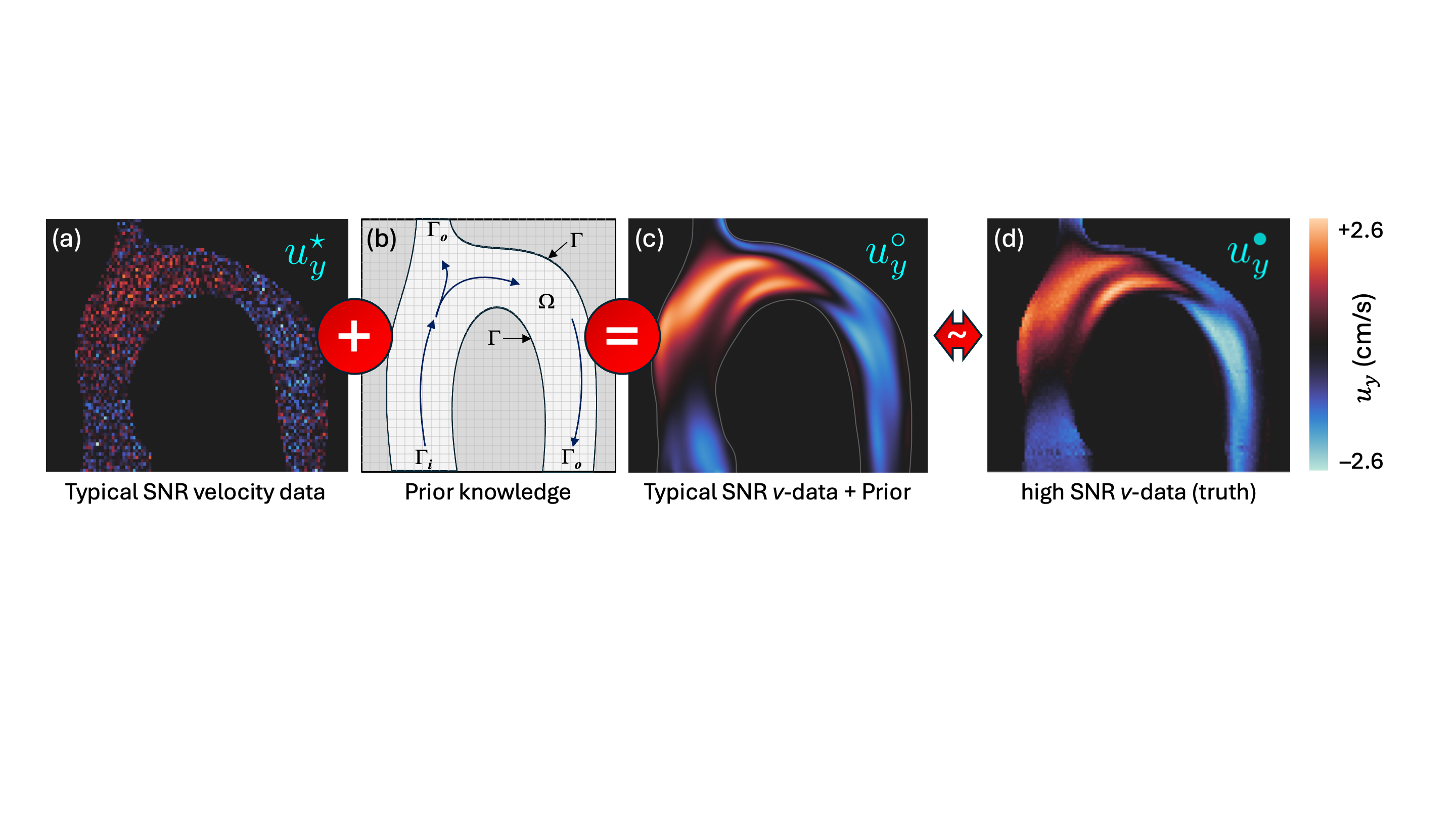}
    \caption{(a) A 2D slice of 3D three-component velocity data from flow MRI through a 3D-printed aorta. This data has typical signal-to-noise-ratio (SNR) and is colored by the velocity into the page, $u_y$. (b) A diagram of our prior knowledge that this velocity data comes from a fluid that obeys the Navier--Stokes equations within a domain $\Omega$ bounded by a no-slip surface $\Gamma$, a Dirichlet velocity condition on $\Gamma_i$, and a stress-free boundary on $\Gamma_o$. This prior knowledge is expressed through a finite element CFD solver in which the free parameters define the shape of $\Gamma$, the velocity on $\Gamma_i$, and the viscosity $\mu$. The optimal parameter values and their uncertainties are found with adjoint-accelerated Bayesian inference giving (c) the inferred velocity field and boundary position, which can be compared with (d) high SNR data gathered over several hours, which was not seen by the assimilation algorithm. The reconstructed field in panel (c) is a higher resolution twin of the high SNR data in panel (d) despite assimilating only the low SNR data in panel (a) and the prior knowledge in panel (b). This figure is adapted from \citet{kontogiannis2024}.}
    \label{fig:FlowMRI_aorta}
\end{figure}

In previous studies, we have demonstrated adjoint-accelerated Bayesian inference on experimental flow MRI data from a laminar flow through a nozzle \citep{Kontogiannis2022} and a 3D-printed aortic arch \citep{kontogiannis2024}. 
Flow MRI data provides three-component velocity fields in a large 3D volume. The spatial resolution is around 1 mm$^3$ per voxel, which is coarse compared with many optical diagnostics, and the signal-to-noise ratio (SNR) is relatively low. Nevertheless, when combined with prior knowledge that this velocity field must obey Navier--Stokes and the no-slip boundary condition, this coarse noisy flow MRI data is sufficient to infer the flow with surprising accuracy. 
For example, Figure \ref{fig:FlowMRI_aorta} shows a reconstruction of 3D three-component flow MRI velocity data of steady flow through a 3D-printed aortic arch at Reynolds number 550. The raw data [Figure \ref{fig:FlowMRI_aorta}(a)] at typical SNR is combined with prior information expressed through a Finite Element Method (FEM), which enforces conservation of mass and momentum [Figure \ref{fig:FlowMRI_aorta}(b)]. The velocity field is inferred rapidly with adjoint-accelerated Bayesian inference [Figure \ref{fig:FlowMRI_aorta}(c)]. The inferred velocity is almost identical to the high SNR velocity data [Figure \ref{fig:FlowMRI_aorta}(d)], which the algorithm did not see. Furthermore, the reconstruction in Figure \ref{fig:FlowMRI_aorta}(c) has higher spatial resolution than the high SNR data and also calculates the position of the aorta’s boundary [thin grey line in Figure \ref{fig:FlowMRI_aorta}(c)]. The other velocity components are presented in \cite{kontogiannis2024}.

The above studies assumed uniform viscosity. In \citet{kontogiannis2024learningrheologicalparametersnonnewtonian}, we measured 3D steady flow MRI data of a shear-thinning jet and assimilated the flow and viscosity model simultaneously. The inferred viscosities were nearly identical to those measured in 18 separate rheometry experiments over four orders of magnitude of shear rate. This showed that it is possible to assimilate velocity data into the Navier--Stokes equations with a non-uniform viscosity model.

This study extends \citet{kontogiannis2024learningrheologicalparametersnonnewtonian} to a turbulent flow from a 10-mm diameter jet at Reynolds number 6500. The spatial resolution of flow MRI is insufficient to resolve turbulent structures in this flow, meaning that turbulence models are required to assimilate this flow MRI data. We use adjoint-accelerated Bayesian inference to combine flow MRI velocity data with prior knowledge that the data are from a turbulent jet that obeys the no-slip condition, and the RANS equations with a prescribed eddy viscosity model. We note that the local turbulent kinetic energy (TKE) can be calculated from the decay rate of the MR signal, although we do not use this information in the current paper. 

\section{Bayesian inversion of RANS problems}
\label{sec:bayesian_inv}
Given noisy flow MRI data of a mean flow, $\bm{u}^\star$, we assume that there is a RANS solution, $\bm{u}^\circ=\bm{u}(\bm{x}^\circ)$, that can fit the data for suitable parameters, $\bm{x}^\circ$, i.e.,
\begin{equation}
\bm{u}^\star = \mathcal{Z}\bm{x}^\circ + \bm{\eps} \quad,
\label{eq:model_approx_data}
\end{equation}
where $\mathcal{Z}\coloneqq\mathcal{S}\mathcal{Q}$ is the operator that maps model parameters to model solutions projected into the data space, $\mathcal{S}$ is the model space to data space projection operator, $\mathcal{Q}$ is the operator that encapsulates the physical model, and $\bm{\eps}\sim\mathcal{N}(\bm{0},\datacov)$ is Gaussian noise with zero mean and covariance operator $\datacov$. We further assume that we have prior knowledge of the probability distribution of the model parameters, $\bm{x}$. We call this the prior parameter distribution, $\mathcal{N}(\mean{\bm{x}},\priorcov)$, which we assume to be Gaussian with prior mean $\mean{\bm{x}}$, and prior covariance operator $\priorcov$. We then use Bayes' theorem, which states that the posterior probability density function (PDF) of $\bm{x}$, given the data $\bm{u}^\star$, $\pi\big(\bm{x}\big|\bm{u}^\star\big)$, is proportional to the data likelihood, $\pi\big(\bm{u}^\star\big|\bm{x}\big)$, times the prior PDF of $\bm{x}$, $\pi\big(\bm{x}\big)$, i.e.,
\begin{align}
\pi\big(\bm{x}\big|\bm{u}^\star\big) &\propto \pi\big(\bm{u}^\star\big|\bm{x}\big)~\pi(\bm{x}) \nonumber\\
&= \exp\Big(-\frac{1}{2}\norm{\bm{u}^\star-\mathcal{Z}\bm{x}}^2_{\datacov} 
- \frac{1}{2}\norm{\bm{x}-\mean{\bm{x}}}^2_{\priorcov}\Big)\label{eq:posterior_pdf_0}\quad,
\end{align}
where $\pi(\cdot)$ is the Gaussian PDF, and $\norm{\cdot,\cdot}^2_{\mathcal{C}} \coloneqq \binner{\cdot,\mathcal{C}^{-1}\cdot}$ is the covariance-weighted \mbox{$L^2$-norm}. The most likely parameters $\bm{x}^\circ$ (maximum \emph{a-posteriori} probability, or MAP estimator), are given implicitly as the solution of the nonlinear optimization problem
\begin{equation}
\quad \bm{x}^\circ \equiv \underset{\bm{x}}{\mathrm{argmin}}\mathscr{J}(\bm{x}) \quad,\quad \text{where}\quad \mathscr{J} \coloneqq -\log\Big(\pi\big(\bm{u}^\star\big|\bm{x}\big)\pi(\bm{x})\Big)\quad.
\label{eq:map_opt_problem}
\end{equation}

For $\bm{x}$ in a neighborhood of the MAP point, $\bm{x}^\circ$, we can {Laplace-approximate} \citep[Chapter~27]{MacKay2003} the posterior distribution such that
\begin{equation}
\pi\big(\bm{x}\big|\bm{y}^\star\big)\simeq  \widetilde{\pi}\big(\bm{x}\big|\bm{y}^\star\big) \coloneqq \exp\Big(-\frac{1}{2}\norm{\bm{x}-\bm{x}^\circ}^2_{\postcovmap} - \textrm{const.}\Big)\quad,
\label{eq:laplace_approx}
\end{equation}
where $\postcovmap \coloneqq \big(\mathcal{G}_{\bm{x}^\circ}^*\datacov^{-1}\mathcal{G}_{\bm{x}^\circ}+\priorcov^{-1}\big)^{-1}$ is the posterior covariance operator, $\mathcal{G}_{\bm{x}^\circ}$ is a linearization of operator $\mathcal{Z}$ around $\bm{x}^\circ$, and $\mathcal{G}^*_{\bm{x}^\circ}$ is the adjoint operator of $\mathcal{G}_{\bm{x}^\circ}$ (see Table \ref{tab:bi_nomen}). The Bayesian inversion of Navier--Stokes problems is treated in more detail in \citet{Kontogiannis2022,kontogiannis2024}.

\begin{table}
\begin{center}
\vskip -0.075in
\begin{tabular}{l l | l l}
\hline
$\mathcal{S}$ & Model-to-data space projection & $\mathcal{Q}$ & RANS problem: maps $\bm{x}$ to $\bm{u}$ \\
$\mathcal{Z}$ & ($\coloneqq \mathcal{S}\mathcal{Q}$) maps $\bm{x}$ to $\mathcal{S}\bm{u}$  & $\mathcal{G}_{\bm{x}_k}$ & Linearization of $\mathcal{Z}$ around $\bm{x}_k$ \\
$\datacov$ & Data covariance operator & $\postcovxk$ & Model parameter covariance operator around $\bm{x}_k$
\end{tabular}
\caption{Operators used in the Bayesian RANS inversion \citep{kontogiannis2024}}
\label{tab:bi_nomen}
\end{center}
\end{table}

\subsection{RANS problem formulation}
\label{sec:gen_ns_problem}
\begin{figure}
\centering
\includegraphics[width=\textwidth]{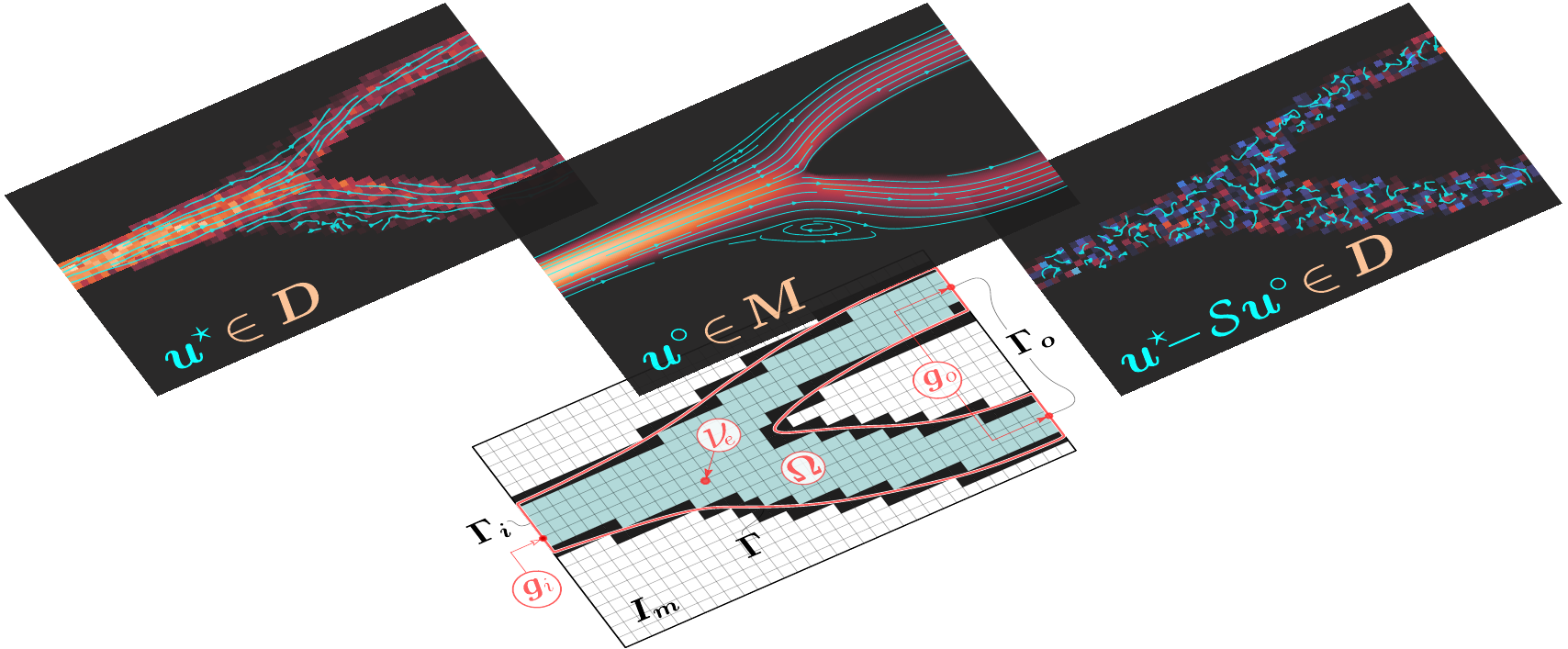}
\caption{Bayesian inversion of RANS problems: we assimilate mean flow data, $\bm{u}^\star$, in order to jointly reconstruct the mean flow field, $\bm{u}$, and learn the unknown RANS parameters, $\bm{x}$. The most likely RANS parameters, $\bm{x}^\circ$, are found by solving problem \eqref{eq:map_opt_problem}, and the inferred mean velocity field is given by $\bm{u}^\circ=\mathcal{Q}\bm{x}^\circ$ \citep{kontogiannis2024}.}
\label{fig:flow_reconstruction_concept}
\end{figure}
The RANS boundary value problem (BVP) in $\Omega \subset I_m \subset \mathbb{R}^3$ (see Figure \ref{fig:flow_reconstruction_concept}) is given by
\begin{equation}
  \left\{\begin{alignedat}{2}
    \bm{u}\bm{\cdot}\nabla\bm{u}-\nabla\bm{\cdot}\big(2{\color{deepred}\nu_e}\nabla^s\bm{u}\big)  + \nabla p &= \bm{0} \quad &&\textrm{in}\quad {\color{deepred}\Omega} \\
    \nabla \bm{\cdot} \bm{u} &= 0 \quad &&\textrm{in}\quad {\color{deepred}\Omega}\\
    \bm{u} &= \bm{0} \quad &&\textrm{on}\quad {\color{deepred}\Gamma} \\
    \bm{u} &= {\color{deepred}\bm{g}_i} \quad &&\textrm{on}\quad {\color{deepred}\Gamma_i} \\
   -2{\color{deepred}\nu_e}\nabla^s\bm{u}\bm{\cdot}\bm{\nu}+p\bm{\nu} &={\color{deepred}\bm{g}_o} \quad &&\textrm{on}\quad {\color{black}\Gamma_o}
  \end{alignedat}\right. \quad,\quad \underbrace{{\color{black}\bm{x}} = ({{\color{deepred}\Omega},{\color{deepred}\bm{g}_i},\color{deepred}\bm{g}_o},{\color{deepred}\nu_e})}_{\text{Unknown RANS parameters}}\quad,
  \label{eq:navierstokes_bvp}
\end{equation}
where $\bm{u}$ is the {mean} velocity, $p\mapsfrom p/\rho$ is the reduced pressure, $\rho$ is the density, $\nu_e$ is the {effective} kinematic viscosity, $\nabla^s\bm{u} \equiv (\nabla^s \bm{u})_{ij} \coloneqq \frac{1}{2}(\partial_j u_i + \partial_i u_j)$ is the strain-rate tensor, $\bm{g}_i$ is the Dirichlet boundary condition (BC) at the inlet $\Gamma_i$, $\bm{g}_o$ is the natural BC at the outlet $\Gamma_o$, and $\bm{\nu}$ is the unit normal vector on the boundary $\partial\Omega = \Gamma\cup\Gamma_i\cup\Gamma_o$, where $\Gamma$ is the part of the boundary where a no-slip BC is imposed. Problem \eqref{eq:navierstokes_bvp} can model many different wall-bounded incompressible laminar flows for Newtonian or generalized Newtonian fluids (i.e., power-law non-Newtonian fluids), as well as turbulent mean flows. Here, since we are interested in modeling turbulent mean flows, we define
\begin{equation}
\mu_e \coloneqq \mu_\ell + \mu_t \quad,\quad \mu_t\coloneqq\ell_c^2\dot{\gamma}\quad, \quad\dot{\gamma}\coloneqq\sqrt{2\nabla^s\bm{u}\bm{:}\nabla^s\bm{u}}\quad,
\end{equation}
where $\mu_e = \rho \nu_e$ is the effective dynamic viscosity, $\mu_\ell$ is the laminar viscosity, $\mu_t$ is the eddy viscosity, $\ell_c$ is a compound mixing length, and $\dot{\gamma}$ is the shear rate magnitude. In this paper, in order to fit the mean flow of a confined turbulent jet, we introduce the compound mixing length
\begin{equation}
\ell = \ell(d_s,d_w;\alpha,\beta,x_c,c,d_{s_0}) \coloneqq \alpha \mathcal{H_\eta}~(d_s-d_{s_0}) + \beta (1-\mathcal{H}_\eta)~d_w\quad,
\label{eq:lc_model}
\end{equation}
where $d_s$ is the distance measured along the flow (streamline) main direction, $d_w$ is the distance from the wall, $\alpha, \beta \in \mathbb{R}$ are the unknown proportionality constants, $d_{s_0}$ is an offset,
\begin{equation}
\mathcal{H}_\eta=\mathcal{H}_\eta(x_c,c) \coloneqq \bigg(1+ \exp{\bigg(\frac{(d_s-x_c)^2-c^2}{\eta c^2}\bigg)}\bigg)^{-1}\quad
\end{equation}
is a smooth activation (Heaviside-like) function, centered around $x_c$, with width $2c$, and $\eta = 0.2$. {Note that, the compound mixing length model that we have introduced is designed for {confined} turbulent jets in order to deal with the two different flow regions: (i) the jet breakdown region, where the mixing length is a function of the streamwise distance, and (ii) the pipe flow region, where the mixing length is a function of the wall distance.} For fixed density $\rho$, $\mu_e$ (or, equivalently, $\nu_e$) is then described by an algebraic six-parameter model such that $\mu_e = \mu_e(\ell_c,\dot{\gamma};\bm{p})$, where $\bm{p}\coloneqq (\mu_\ell,\alpha,\beta,x_c,c,d_{s_0})$. The unknown RANS parameters thus become $\bm{x}=(\Omega,\bm{g}_i,\bm{g}_o,\bm{p})$.

\section{The flow MRI experiment}
\label{sec:flow_mri_exp}

\subsection{Phantom design}
\begin{figure}
    \centering
    \includegraphics[width=0.8\linewidth]{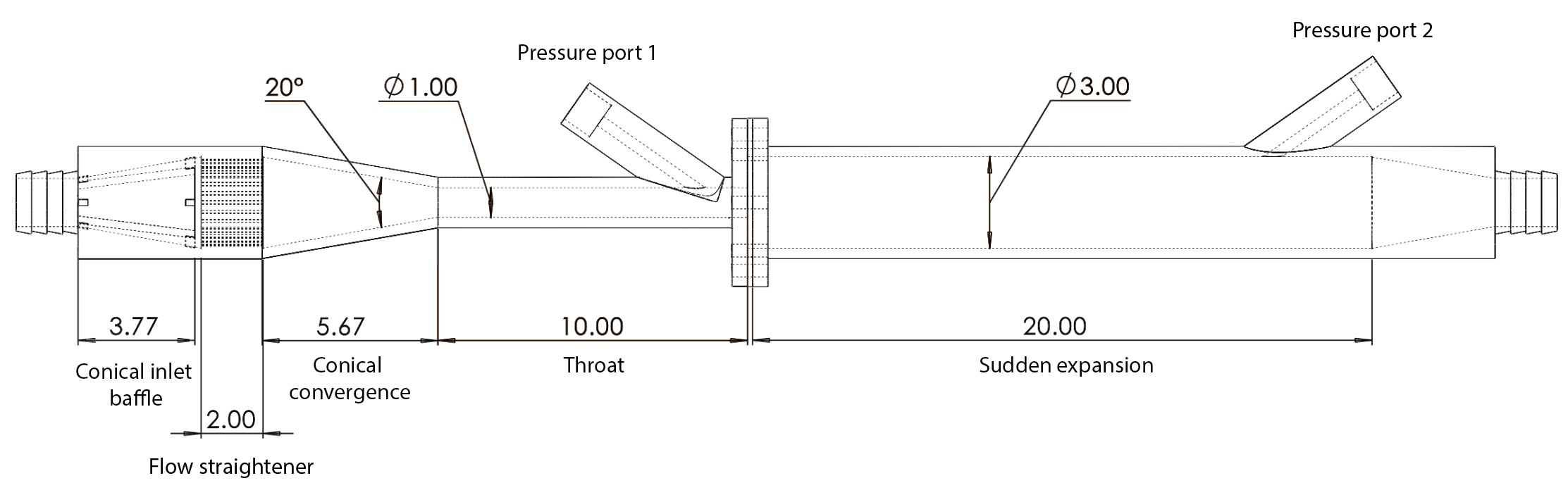}
    \caption{Phantom design schematic (length units in centimeters).}
    \label{fig:phantom_design_fda}
\end{figure}

The phantom was designed in SolidWorks based on the FDA nozzle geometry definition with modifications to fully develop the inlet flow and accommodate direct pressure measurements. The nozzle design, which is shown in Figure \ref{fig:phantom_design_fda}, includes the following main features:
\begin{enumerate}
    \item[(1)] \emph{Conical inlet baffle}. A conical inlet baffle upstream of the inlet expansion zone was added to reduce the opening angle and thereby prevent formation of a jet. The 1:1 ratio of the areas of the two paths at the inlet was maintained at the outlet to the baffle.
    \item[(2)] \emph{Flow straightener}. A grid-like flow straightener was added prior to the converging section to avoid the introduction of jets in the region of interest. The structured grid has square channels 0.2$\times$0.2 cm with 0.1-cm walls between them.
    \item[(3)] \emph{Conical convergence}. A convergence zone of 20$^\circ$ brings the diameter of the nozzle down to the diameter of the throat.
    \item[(4)] \emph{Throat}. A narrow segment with a diameter of 1 cm and a length of 10 cm (ten times the diameter).
    \item[(5)] \emph{Sudden expansion}. The diameter of the nozzle expands to 3 cm (three times the diameter of the throat) resulting in creation of a jet at sufficient Reynolds numbers.
    \item[(6)] \emph{Pressure ports}. Two Luer lock ports were added. One port is proximal to the sudden expansion and one port is downstream (after the jet has dissipated). Pressure catheters were inserted through these ports to acquire estimates of pressure drop between the two points. 
\end{enumerate}

\begin{figure}
\centering
\begin{subfigure}{.6\textwidth}
    \centering
    \includegraphics[width=\linewidth]{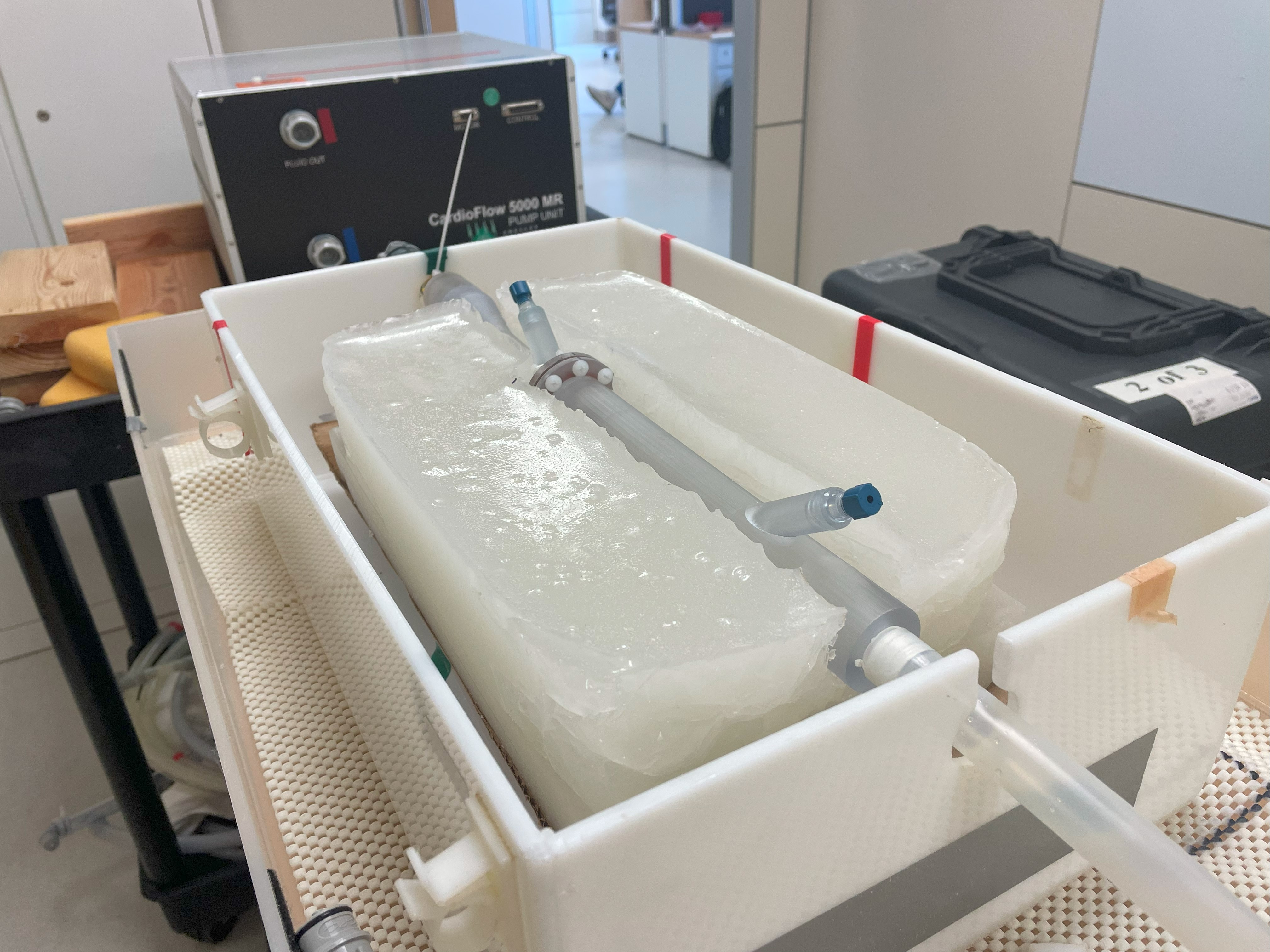}
    \caption{}
    \label{fig:enter-label}
\end{subfigure}
\begin{subfigure}{.625\textwidth}
    \centering
    \includegraphics[width=\linewidth]{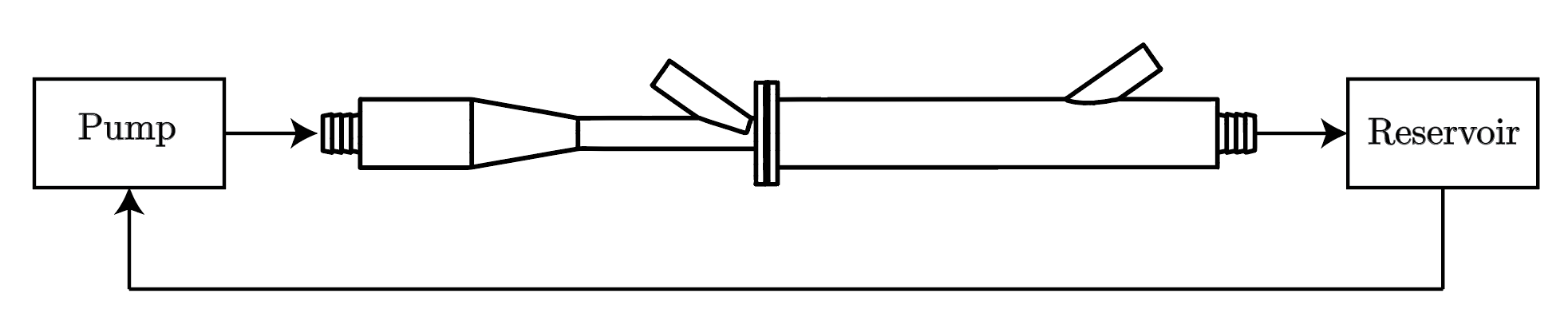}
    \caption{}
    \label{fig:enter-label}
\end{subfigure}
\caption{Nozzle phantom assembled on the benchtop. (a) Assembled phantom embedded into a solid ballistics gel. (b) Bench setup schematic.}
\label{fig:benchtop}
\end{figure}

\subsection{Flow bench setup}
Given restrictions on the printer test bed size, the phantom design was split into two parts right after the throat (see Figure \ref{fig:benchtop}). The two pieces were printed individually using Clear Loctite 3D IND405 in a Carbon L1 3D printer to have rigid walls with a wall thickness of 0.3 cm. The two parts were screwed together with a rubber gasket in between to ensure the interface was watertight. The assembled phantom was embedded into a solid ballistics gel (ClearBallistics) for spatial stabilization and to provide static MRI signal regions for velocity (phase) offset corrections. An MRI-compatible programmable flow pump (CardioFlow 5000 MR, Shelley Medical Imaging Technologies) was used to pump flow through the phantom at a steady flow rate of 200 mL/s, corresponding to a Reynolds number of 6500. A glycerol-water mixture (40/60 ratio) was used as a blood-mimicking fluid. Ferumoxytol, a contrast medium, was added for MRI signal enhancement. The density of the fluid was 1.1 g/cm\textsuperscript{3} and the dynamic viscosity was 4.2 mPa.s. Estimates of pressure drop were acquired using two pressure transducers (SPR-350S, Millar), one inserted proximal to the sudden expansion and another further downstream after the jet had dissipated. MRI data was acquired on a 3T scanner (Siemens Vida Fit) in the phantom with a conventional 4D-flow measurement sequence using an 18-channel chest coil. The scan was acquired with a resolution of 1$\times$1$\times$1 mm\textsuperscript{3} (isotropic), a matrix size of 256$\times$256$\times$44, TE/TR = 3.00/7.14 ms, a flip angle of 7°, GRAPPA R = 2, and an acquisition time of 4 minutes.  Four different data sets were acquired with velocity encoding strength (VENC) of [500, 200, 100, 50] cm/s. The final imaging data set was generated as a magnitude weighted average of the different VENC acquisitions, unwrapped against the \mbox{VENC=500 cm/s} data set.


\section{Reconstruction of flow MRI data}
We apply the RANS inversion methodology [see section \ref{sec:bayesian_inv} and \cite{kontogiannis2024}\footnote{For the Bayesian RANS inversion algorithm, we use the same numerical stabilization parameters as in \cite{kontogiannis2024}.}] to reconstruct 3D mean flow data of a confined turbulent jet measured using flow MRI (see section \ref{sec:flow_mri_exp}). We assume that the noise in the measured velocity field is white Gaussian \citep{Gudbjartsson1995}, with zero mean and covariance operator $\mathcal{C}_{\bm{u}^\star} = \sigma\mathrm{I}$, where $\sigma = 5$ cm/s. We fix the geometry, $\Omega$, which is found by segmenting the magnitude image of the flow MRI scan, and set the outlet BC to $\bm{g}_o\equiv\bm{0}$. We thus only need infer the unknowns $\bm{x}=(\bm{g}_i,\bm{p})$ in order to reconstruct the mean flow data, $\bm{u}^\star$. The dimensionality of the data and model spaces is given in Table \ref{tab:data_and_model_spaces}.

\begin{figure}
    \centering
    \includegraphics[width=0.7\linewidth]{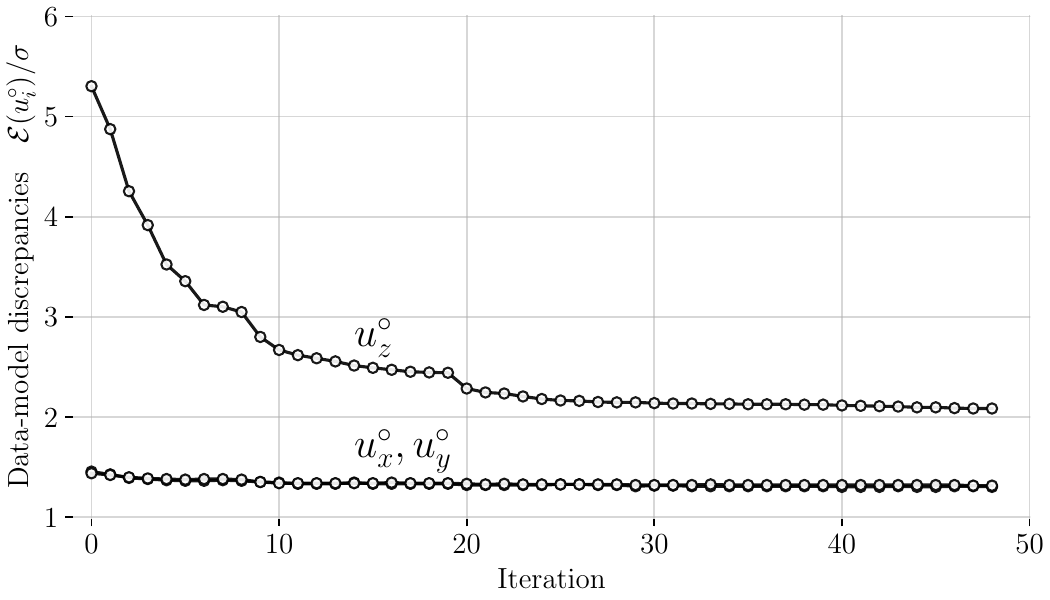}
    \caption{Optimization log of the data-model discrepancy of each velocity component.}
    \label{fig:opt_log}
\end{figure}

We define the component-wise data-model discrepancy by 
\begin{equation}
\mathcal{E}({u}_i) \coloneqq \abs{\Omega}^{-1}\norm{u_i^\star-\mathcal{S}u_i}_{L^2(\Omega)}\quad.
\end{equation} 
We find that the prior-data discrepancies are $\big(\mathcal{E}(\bar{u}_x),\mathcal{E}(\bar{u}_y),\mathcal{E}(\bar{u}_y)\big)/\sigma=\big(1.45,1.44,5.30\big)$, and the MAP-data discrepancies are $\big(\mathcal{E}({u}^\circ_x),\mathcal{E}(\bar{u}^\circ_y),\mathcal{E}(\bar{u}^\circ_y)\big)/\sigma=\big(1.30,1.31,2.08\big)$. 
The Bayesian RANS inversion algorithm takes 48 iterations to converge; 95\% of the error reduction, however, is achieved in the first 20 iterations (see Figure \ref{fig:opt_log}).
The RANS model reconstruction (i.e., the inferred mean flow) and the discrepancies are shown in Figure \ref{fig:vel_rec_err}(f). The inversion algorithm infers the velocity inlet BC, $\bm{g}_i$, and the effective viscosity field parameters, $\bm{p}$, such that the discrepancies between the RANS model and the flow MRI data are minimized. The inferred algebraic viscosity model parameters, as well as their uncertainties, are shown in Table \ref{tab:lc_parameters}.
The prior and the learned (MAP) effective viscosity fields, which are functions of the parameters $\bm{p}$, are shown in Figure \ref{fig:eff_visc_learning}. 
\subsection{Discussion}
\begin{figure}
\centering
\begin{subfigure}{.49\textwidth}
\begin{center}
	    	\reflectbox{\rotatebox[origin=c]{180}{\includegraphics[width = 1\textwidth]{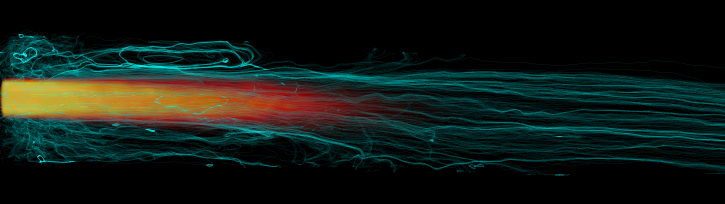}}}
      	\caption{}
       \label{subfig:flowmri_stream}
\end{center}
\end{subfigure}
\centering
\begin{subfigure}{.49\textwidth}
\begin{center}
	    	\reflectbox{\rotatebox[origin=c]{180}{\includegraphics[width = 1\textwidth]{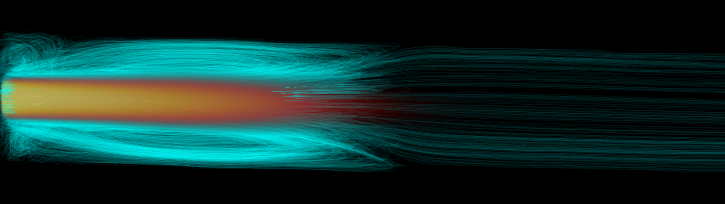}}}
      	\caption{}
       \label{subfig:rec_stream}
\end{center}
\end{subfigure}
\begin{subfigure}{.49\textwidth}
\begin{center}
	    \includegraphics[width = 1\textwidth]{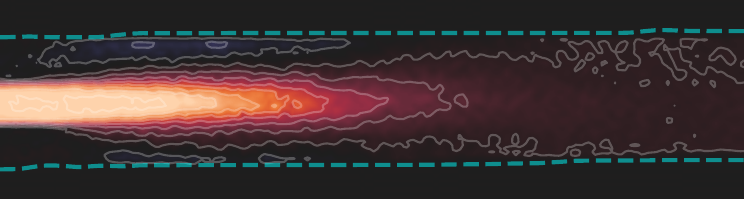}
      	\caption{}
       \label{subfig:flowmri_uz}
\end{center}
\end{subfigure}
\begin{subfigure}{.49\textwidth}
\begin{center}
	    	\includegraphics[width = 1\textwidth]{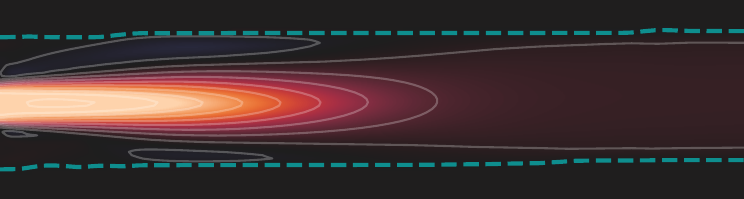}
      	\caption{}
       \label{subfig:rec_uz}
\end{center}
\end{subfigure}
\begin{subfigure}{.49\textwidth}
\begin{center}
	    	\includegraphics[width = 1\textwidth]{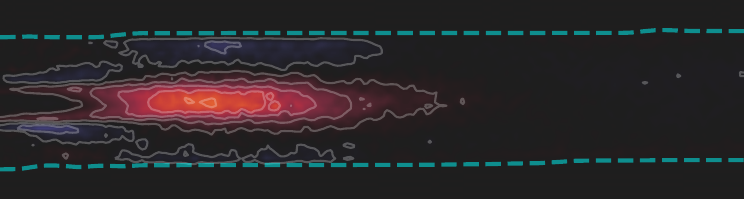}
      	\caption{}
       \label{subfig:prior_err}
\end{center}
\end{subfigure}
\begin{subfigure}{.49\textwidth}
\begin{center}
	    	\includegraphics[width = 1\textwidth]{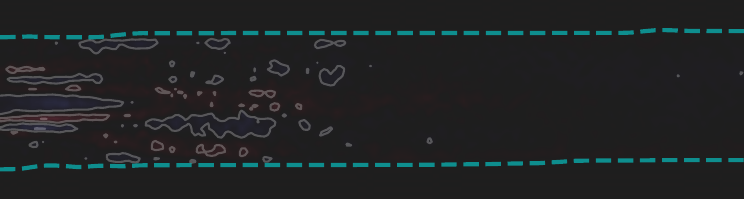}
      	\caption{}
       \label{subfig:rec_err}
\end{center}
\end{subfigure}
\caption{Noisy 3D flow MRI data, $\bm{u}^\star$, and RANS model reconstruction, $\bm{u}^\circ$. (a) Flow MRI data, $\bm{u}^\star$. (b) Inferred (MAP), ${u}_z^\circ$. Images (a,b) show velocity streamlines in cyan color and velocity magnitude in red color. (c) Flow MRI data, ${u}_z^\star$. (d) Inferred (MAP), ${u}_z^\circ$. (c,d) The mid-plane axial velocity ($u_z$) slice is shown and the dashed cyan line depicts the walls of the confined turbulent jet. (e) Data-prior discrepancy, $\mathcal{S}^*({u}_z^\star-\mathcal{S}\bar{u}_z)$. (f) Data-MAP discrepancy, $\mathcal{S}^*({u}_z^\star-\mathcal{S}{u}_z^\circ)$. (e,f) The data-model discrepancy is shown before and after assimilation of the flow MRI data. Panels (c--f) share the same colormap scale, which is [0,100cm/s].}
\label{fig:vel_rec_err}
\end{figure}

\begin{figure}
\centering
\begin{subfigure}{.49\textwidth}
\begin{center}
	    	\includegraphics[width = 1\textwidth]{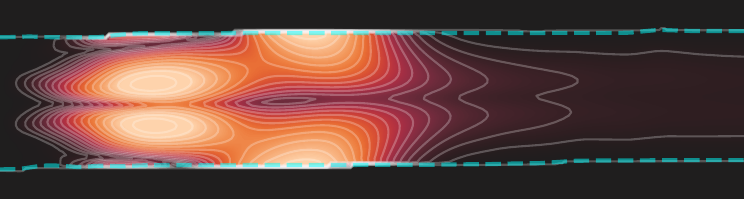}
      	\caption{}
       \label{subfig:visc_prior}
\end{center}
\end{subfigure}
\begin{subfigure}{.49\textwidth}
\begin{center}
	    	\includegraphics[width = 1\textwidth]{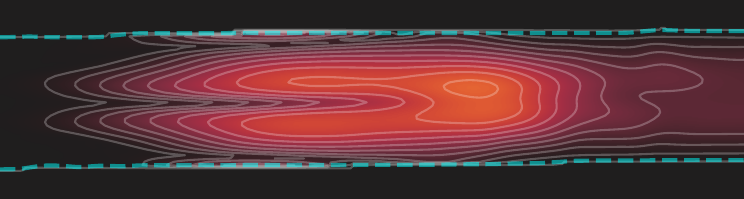}
      	\caption{}
       \label{subfig:visc_rec}
\end{center}
\end{subfigure}
\begin{subfigure}{.49\textwidth}
\begin{center}
	    	\includegraphics[width = 1\textwidth]{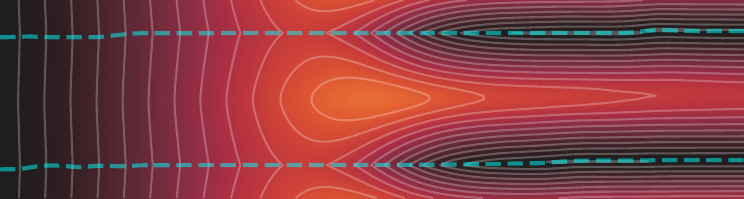}
      	\caption{}
       \label{subfig:lc_prior}
\end{center}
\end{subfigure}
\begin{subfigure}{.49\textwidth}
\begin{center}
	    	\includegraphics[width = 1\textwidth]{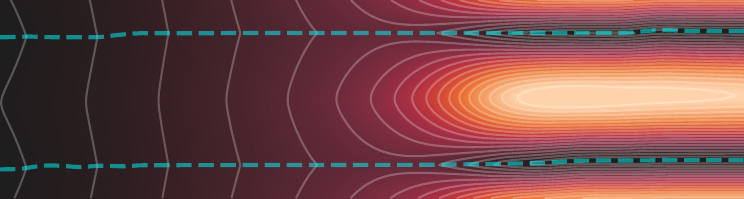}
      	\caption{}
       \label{subfig:lc_rec}
\end{center}
\end{subfigure}
\begin{subfigure}{.49\textwidth}
\begin{center}
	    	\includegraphics[width = 1\textwidth]{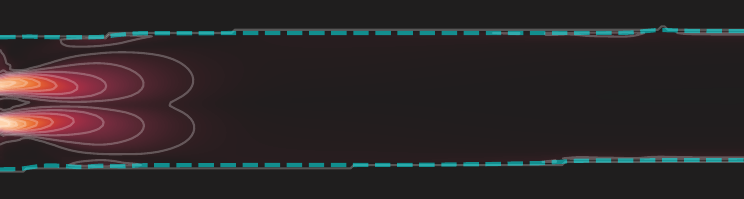}
      	\caption{}
       \label{subfig:gamma_prior}
\end{center}
\end{subfigure}
\begin{subfigure}{.49\textwidth}
\begin{center}
	    	\includegraphics[width = 1\textwidth]{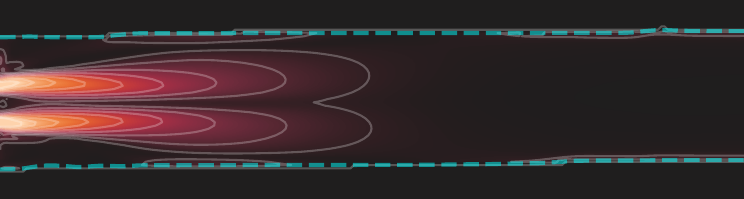}
      	\caption{}
       \label{subfig:gamma_rec}
\end{center}
\end{subfigure}
\caption{(a,c,e) Prior assumption and (b,d,f) inferred MAP estimate for (a,b) the effective viscosity field, $\mu_e \coloneqq \mu_\ell + \ell^2_c\dot{\gamma}$, (c,d) the compound mixing length, $\ell_c$, and (e,f) the shear rate magnitude, $\dot{\gamma}(\bm{u})$ (the dashed cyan line depicts the walls of the confined turbulent jet). Panels (a,b), (c,d), and (e,f) share the same colormap scale.}
\label{fig:eff_visc_learning}
\end{figure}

In this paper we use a five-parameter algebraic turbulence model for the effective viscosity field (Boussinesq approximation), which is based on a compound mixing length, such that $\mu_t = \ell^2_c\dot{\gamma}$. 
{Figure \ref{fig:vel_rec_err}(f) shows that the algorithm has managed to find model parameters that cause the inferred velocity field to closely match the measured velocity field. To do so, it has significantly altered the effective viscosity and compound mixing length fields from their prior values, as shown in Figure \ref{fig:eff_visc_learning}. In particular, the RANS inversion algorithm learns the compound mixing length parameters that generate a modeled velocity field that fits both the jet breakdown region, and the pipe flow region. This shows that this RANS model is sufficiently descriptive for this confined jet flow.}
{These results prove the concept of assimilating RANS model parameters from flow MRI data.} 

{Future investigations will consider different turbulence models.}
A different, dimensionally consistent algebraic model based on the turbulence kinetic energy, $k$, is given by $\mu_t = \ell_ck^{1/2}$. This model requires TKE data, and has been used in \cite{CASEL2022111695} yielding good results. 
{This will be examined next; however,} it is important to note that algebraic models based on a learned compound mixing length and TKE data can be made to fit the measured mean flow but are not expected to extrapolate well at different flow conditions. 
{A further improvement will be obtained from} one- or two-equation turbulence models, such as the $k$-$\epsilon$ turbulence model, because it should extrapolate well outside the training regime.

\begin{table}
\small
  \begin{center}
\centering
  \begin{tabular}{ccc|ccc}
        \multicolumn{2}{c}{Data voxels}   &  {Data resolution} &  \multicolumn{2}{c}{Model voxels} & {Model resolution}\\[3pt]
        \multicolumn{2}{c}{$40\times45\times169$} & {$1\times1\times1$ mm}  & \multicolumn{2}{c}{$53\times60\times225$} & {$0.75\times0.75\times0.75$ mm}\\[6pt] 
  \end{tabular}\\
  \caption{Data and model spaces for the Bayesian inverse RANS problem.}
  \label{tab:data_and_model_spaces}
  \end{center}
\end{table}

\begin{table}
\begin{center}
\vskip -0.075in
\begin{tabular}{c c c c c c}
\multicolumn{6}{c}{Inferred (MAP estimate)}\\[-0.25em]
\hline
$\mu_\ell$ [mPa.s] & $\alpha$ & $\beta$  & $x_c$ [cm] & $c$ [cm] & $d_{s_o}$ [cm]\\
\vspace{0.1in}
4.2 $\pm$ 0.15 & 0.722 $\pm$ 0.193 & 19.2 $\pm$ 12.6 & 3.25 $\pm$ 2.87 & 7.32 $\pm$ 2.44 & 0.159 $\pm$ 0.0881\\
\multicolumn{6}{c}{Prior}\\[-0.25em]
\hline
$\mu_\ell$ [mPa.s] & $\alpha$ & $\beta$ & $x_c$ [cm] & $c$ [cm] & $d_{s_o}$ [cm]\\
4.2 $\pm$ 0.15 & 2 $\pm$ 6 & 10 $\pm$ 15  & 0 $\pm$ 30 & 8.5 $\pm$ 30 & 0 $\pm$ 15
\end{tabular}
\caption{Compound mixing length parameters.\label{tab:lc_parameters}}
\end{center}
\end{table}


\section{Conclusions}
We have solved a Bayesian inverse Reynolds-averaged Navier--Stokes (RANS) problem that assimilates 3D velocimetry data of a mean turbulent flow in order to jointly reconstruct the flow field and learn the unknown RANS parameters. By incorporating an algebraic effective viscosity model into the RANS
problem, we devise an algorithm that learns the most likely viscosity model parameters, and estimates their uncertainties, from velocimetry data. Then we conduct a
flow MRI experiment to obtain 3D velocimetry data of a confined turbulent jet through
an idealized medical device, known as the Food and Drug Administration (FDA) nozzle. We show that the algorithm successfully reconstructs the noisy flow field, and, at the same time, learns the
viscosity model parameters and their uncertainties. In this paper we have considered a five-parameter algebraic viscosity model. The present methodology, however, accepts any type of turbulence model, be it algebraic, one-equation, or multi-equation, as
long as the model is differentiable. More complicated turbulence models,
such as the $k$-$\epsilon$ model, can be used in order to reconstruct the flow and learn a turbulence model that can better extrapolate at different flow conditions. The turbulence model parameters can be learned by assimilating velocimetry data into the RANS problem and turbulent kinetic energy (TKE) data into the turbulence model of the kinetic energy equation. {This pilot study has proven the concept. Our next step will be to assimilate more complicated turbulence models.}

\bibliographystyle{ctr}
\bibliography{brief}

\end{document}